# Domain wall propagation in magnetic nanowires by spin polarized current injection

Nicolas Vernier[*], Dan A. Allwood, Del Atkinson, Michael D. Cooke,

Russell P. Cowburn

Nanomagnetism Group, Department of Physics, University of Durham, Rochester Building, Science Laboratories, South Road, Durham DH1 3LE, U.K.

**Abstract**

We demonstrate movement of a head-to-head domain wall through a magnetic nanowire simply by passing an electrical current through the domain wall and without any external magnetic field applied. The effect depends on the sense and magnitude of the electrical current and allows direct propagation of domain walls through complex nanowire shapes, contrary to the case of magnetic field induced propagation. The efficiency of this mechanism has been evaluated and the effective force acting on the wall has been found equal to $0.88 \times 10^{-9}$ N.A$^{-1}$.

PACS: 72.25.Pn, 75.47.Np, 75.60.Cu, 85.75.-d

[*] On leave from and now at Laboratoire de Physique des Solides, bât. 510, Université Paris-Sud, 91405 Orsay cedex, France. Corresponding author, e-mail : vernier@lps.u-psud.fr

Controlling magnetization directly with an electric current rather than a magnetic field is one of the recent exciting developments within spintronics. The main expectations are a very fast reversal of magnetization (< 100 ps) and the ability to control individual magnetic elements without affecting neighbouring structures. In ferromagnetic/non-magnetic/ferromagnetic multilayer structures, several experiments have demonstrated magnetization reversal of one layer purely by applying a pulsed current and without any externally applied magnetic field [1]. This concept could be also very interesting in the framework of recently demonstrated magnetic logic [2], where logic operations are performed by domain wall propagation through ferromagnetic nanowire junctions, resulting in magnetization reversal. Until now, this propagation has been induced by an external applied field. A potential alternative is current-induced dragging of a domain wall, which does not rely on a generated magnetic field and is now well known in quasi-infinite thin film [3,4]. However, the effect of nanopatterning has never been studied, despite introducing several new conditions and raising important questions. First, the shape anisotropy of certain nanowires (e.g. 5 nm thick, 100 – 200 nm wide permalloy) dictates that its magnetization is parallel to the wire axis, and domain walls become head-to-head rather than the Bloch walls studied previously. Second, the profile of the domain wall is modified; its width has been shown to be much smaller in nanostructures [5]. Third, the edge roughness of nanowires could become very important, possibly severely impairing domain wall propagation. Fourth, can domain walls propagate through nanowires that are not straight?

Here, we address these questions and demonstrate experimentally that domain walls can be moved through magnetic nanowire circuits and even around corners by spin polarized current injection. No magnetic field is required and the direction of domain wall motion is defined by the current direction only.

Experimental nanostructures consisting of a U-shaped planar magnetic nanowire connected to non-magnetic electrical contact pads (Fig. 1) were fabricated on oxidized silicon (oxide thickness = 0.6 µm) by two-step electron-beam lithography. The magnetic U-shape had a wire width of 120 nm and was made from 5 nm thick thermally-evaporated Permalloy ($Ni_{81}Fe_{19}$). A 600 nm × 3 µm magnetic 'nucleation pad' at one end of the nanowire was used to inject domain walls into the wire at magnetic fields lower than the usual wire coercivity [6]. The electrical contact pads allowed an electrical current to be passed through the nanowire – the sign convention used in this paper is that positive current corresponds to electrons flowing from position P to position U in Fig. 1B. In-plane alternating magnetic fields $H_x$ and $H_y$ were applied in the $x$ and $y$ directions respectively (see Fig. 1B) with a 90° mutual phase shift so that the applied field rotated in the plane of the sample at 27 Hz. Wire magnetization was measured at position T at room temperature using a magneto-optical Kerr effect (MOKE) magnetometer with a focused laser probe of 5 µm diameter [7]. The magnetometer was configured to measure the x-component of magnetization.

The shape of the nanostructure was carefully designed to separate domain wall nucleation and propagation. This is of central importance, allowing investigation of the effect of electrical current on the propagation process only. By applying a counter-clockwise rotating magnetic field, a domain wall can be injected from the pad and propagated around the different parts of the structure before being annihilated at the 'open' end of the nanowire beneath contact pad 'U' (Fig. 1). In all of the experiments and field-sequences described in this paper, the domain wall always originates from the nucleation pad at the top right hand side and never from the open end of the wire. Crucially, placing the focused laser probe close to point P (Fig. 1) yields information about domain wall nucleation and injection, while placing the probe close to point T (Fig. 1) yields information about the domain wall propagation field: the two are decoupled. For example, when we apply a counter-clockwise rotating field with $H_x$ amplitude of 95 Oe and $H_y$ amplitude of 47 Oe and place the focused magnetooptical laser probe close to point P, we observe that domain walls are injected from the nucleation pad into the nanowire when $H_x$ passes through the value of 40 Oe. This is the domain wall injection field. However, if we now move the laser probe to point T, we measure switching as $H_x$ passes through the much lower value of 20 Oe. This value is the domain wall propagation field, $H_p$, since switching follows immediately from domain wall propagation through the wire section R and corner S (Fig. 1) and not from a nucleation event. The same experiment performed with a clockwise rotating field leads to switching being observed at point P, but no switching at point T, because the sense of field rotation is not matched to the turning sense of corners Q and S, as is consistent with our claim that domain walls always begin from the nucleation pad.

In order now to measure the influence of an electrical current on the domain wall propagation field, we measured magnetization hysteresis loops from point T while applying a counter-clockwise rotating magnetic field ($H_x$ amplitude = 112 Oe and $H_y$ amplitude = 53 Oe) and applying d.c. current, $I$, through the nanostructure. We found that the loops remained symmetrical for all $I$, allowing $H_p$ to be measured and plotted against the current density deduced from $I$ (Fig. 2A). $H_p$ is seen to decrease with increasing current magnitude $|I|$ but importantly, $H_p$ is always lower for $I < 0$ than for $I > 0$. Fig. 3 shows examples of two of the hysteresis loops behind these data. We measure the dependence of the propagation field on the sign of the electrical current through $\Delta H_p = H_p(-I) - H_p(+I)$. This is a measure of the force exerted on the domain wall by the current and is seen to increase monotonically as a function of $|I|$ (Fig. 2B). The force acts in the same direction as the electron flow, regardless of whether the domain wall is a head-to-head or tail-to-tail type.

To demonstrate directly the movement of a domain wall under the action of only a spin polarized current without any magnetic field at all, we replaced the rotating field with a more complex sequence of field pulses (Fig. 4A). A current of either +350 µA or -350 µA was passed through the nanowire throughout the entire experiment. At the start of each sequence a state of continuous magnetization was achieved by applying x and y-directed field pulses up to time t = 0.040 s (Fig. 4A). A single domain wall was then injected from the nucleation pad by a negative *x*-directed pulse at time t = 0.053 s. This field pulse propagates the the wall to position Q but no further as the following vertical section of the wire is orthogonal to the applied field pulse. The applied fields were then switched to zero at $t$ = 0.063 s (Fig. 4B). Magnetooptical magnetization measurements were made as a function of time with the laser probe at point T (Fig. 1). The sequence was repeated every 80 ms for approximately 30 minutes to reduce random noise.

Figure 4B shows the magnetization of the nanowire's lower horizontal arm during application of this field pulse sequence and $I = \pm 350$ µA. With $I = -350$ µA, no magnetization reversal is observed [*8*]. In this case, electrons are flowing from U to P, and so are unable to help propagate the domain wall from its starting position Q to the probing position T. However, for $I = +350$ µA, we observe full magnetization reversal between $t = 0.063$ s and $t = 0.075$ s (Fig. 4B), *even though zero magnetic field is being applied to the sample*, and the nearest domain wall was placed 70 µm away at position Q. There is no possibility of this reversal being a delayed reaction to the field pulse launched at $t = 0.053$ s, as this was oppositely directed. To achieve this reversal, the spin polarized current must have pushed the domain wall all the way from point Q through to point T. As further confirmation that this is indeed what is happening, one also sees a sharp transition at $t = 0.026$ s during the 'reset' phase of the field sequence. Although this transition is due to applied fields, its existence confirms current-induced domain wall propagation since there would be no reversed magnetization to be reset if the initial magnetization state had not been switched by movement of the wall by the spin polarised current [9].

Figure 2 suggests that two distinct mechanisms are at play when an electrical current is passed through a domain wall. The first is a decrease in $H_p$ with increasing $|I|$, which does not depend upon the sign of the current. Most of the variation of $H_p$ with $I$ in Fig. 2A is due to this mechanism. The second is a change in $H_p$ which is sensitive to both the direction and magnitude of current, and which is quantified through $\Delta H_p$.

The sign-independent mechanism can most probably be attributed to thermal activation due to ohmic heating of the magnetic nanowire. A current of 350 µA in these nanowires corresponds to a very high current density of $6 \times 10^{11}$ Am$^{-2}$ and some heating is inevitable. However, resistance measurements suggest a maximum temperature rise of ~100°C and the change in MOKE signal upon magnetization reversal was approximately constant throughout measurements, indicating little change in magnetic moment or temperature relative to the 600°C Curie temperature of Permalloy [*10*]. The current-induced transition in fig. 4B is quite broad, which is a further indication of thermally-activated de-pinning [*11*, *12*]. The process of lowering $H_p$ clearly requires further experiment to be better understood. It should be appreciated that were it not for this lowering of the propagation field, the sign-dependent effect of the spin polarized current in this experiment would not have been strong enough to move a domain wall on its own.

The sign-dependent effect of current provides a force for assisting or hindering domain wall propagation and has a maximum observed value of $\Delta H_p$ = 3.0 Oe. Fig. 2B suggests that this force increases linearly with the current, however, one should be careful as the temperature of the nanowire was also current dependent. A common difficulty in studying the magnetic effects of spin polarized current is to distinguish between the effect due to the classical magnetic field associated with the current and the effect due to the spin polarisation of the carriers. We believe that none of the observed dependence of $\Delta H_p$ on $I$ in our experiment is due to classical magnetic effects. Indeed, we always observe hysteresis loops that are symmetric about zero field and so the interaction between the current and the domain wall must be independent of whether the wall is a positively-charged head-to-head wall or a negatively-charged tail-to-tail wall. A classical magnetic field could shift a hysteresis loop on the field axis, but could not make a loop narrower or broader.

Second, classical 'hydrodynamic drag', as produced by current inhomogeneities close to the wall needs films thicker than 100 nm to be really efficient [*13*, 4]. Furthermore, no current inhomogeneities are expected here as the Lorentz force is zero.

Indeed, because of the shape anisotropy, the magnetization lies along the wire axis and is parallel to the electron velocity, so the Lorentz force **V**x**M** reduces to zero.

The explanation which we find most convincing is that of s-d coupling, first suggested by Berger [*14*]. Recently reported work on current-induced magnetization reversal of multi-layers structures [*15*] is also relevant. The effect is due to an angular momentum transfer from the conduction electrons going through the domain wall. The conduction electrons being mainly in the non-polarized s-band, the momentum arises from a coupling with the spin polarized d-band. This mechanism has already been claimed as the origin of the dragging of Bloch domain walls in thin permalloy films [4]. The fact that Reference [14] assumed a Bloch wall is not a problem as the transposition to head-to-head walls is straightforward. From the fit with a straight line on fig. 2B, we can estimate the pressure on the wall due to the current. $\Delta H_p$ (Fig 2B), although coming from spin polarised current injection, can be considered as an effective applied magnetic field. The pressure exerted on a domain wall by an applied field H is given by the well known formular $2M_SH$ [*16*], where $M_S$ is the saturation magnetization of the magnetic material (860 emu cm$^{-3}$ for permalloy). Thus, for the maximum current density used in this study (700 GA.m$^{-2}$) the spin transfer effect applies a pressure of 515 Pa to the domain wall. In order to make comparisons with other experimental geometries, it is helpful to express this as a pressure per unit current density, which can be simplied to the dimensions of force per unit current. In our experiment, we find a force per unit current of $0.88 \times 10^{-9}$ N.A$^{-1}$. This value compares well with that of $0.6 \times 10^{-9}$ N.A$^{-1}$ found for Bloch walls in previous experiments [4]. A perfect agreement would not be expected as the experimental conditions were very different: the type of domain wall as well as the anisotropy were different.

In conclusion, the experiment presented here shows that it is possible to drag a magnetic domain wall in patterned magnetic nanowires with an electrical current without the assistance of an applied magnetic field. This result is consistent with the s-d coupling theory of Berger, which can apply both to Bloch walls and to the head-to-head walls used in this study. We have obtained quantitative values for the pressure due to this current dragging. These results allow an alternative means of interfacing between electronic and magnetic systems in nanoscale devices and so are potentially very important for applications such as magnetic memory, magnetic logic and spintronics.

# Figure captions

Figure 1. (A) Electron micrograph of a U-shaped magnetic nanowire beneath nonmagnetic electrical contact pads. Fabrication was by electron-beam lithography using a 30 kV electron acceleration voltage, polymethylmethacrylate resist, metalization by thermal evaporation and performing lift-off in acetone. The nanostructure had 80 μm long horizontal arms, 60 μm long vertical arm and corners with a turning radius of 10 μm. The inset shows a high magnification image of a vertical part of the nanowire. (B) Schematic diagram of U-shaped nanowire and electrical contact pads and definition of *x*- and *y*-directions.

Figure 2. (A) Horizontal magnetic field for domain wall propagation, $H_p$, of lower arm of the magnetic nanowire in a counter-clockwise applied rotating magnetic field having $H_x = 112$ Oe and $H_y = 53$ Oe, as a function of the magnitude of the current passing through the nanowire. ■ show negative current, ● show positive current. (B) the difference, $\Delta H_p$, between $H_p$ values for positive and negative current [$\Delta H_p = H_p(-I) - H_p(+I)$], as a function of current magnitude.

Figure 3. Magnetic hysteresis loops obtained using a counter-clockwise applied rotating magnetic field with $H_x = 112$ Oe and $H_y = 53$ Oe from the lower horizontal arm for positive current (dashed line) and negative current (solid line). The absolute value of the intensity |*I*| was 325 μA.

Figure 4. (A) Sequence of horizontal and vertical magnetic field pulses applied to the magnetic nanowire. $H_x$ = solid line, $H_y$ = dashed line. The pulses up to t = 0.040 s reset the magnetization of the nanowire to a continuous state; the pulse at t = 0.052 s then injects a single domain wall which moves as far as point Q (Fig. 1B). (B) the magnetooptical signal obtained during such a sequence from the nanowire's lower horizontal arm with current, $I$ = -350 µA (dashed line) and $I$ = +350 µA (solid line). The first transition in the positive current trace at $t$ = 0.027 s is the system being reset under the action of the applied field; the second transition at $t$ = 0.065 s is the domain wall moving through point T (Fig. 1B) under the action of only the spin polarized current.

**Figure 1**

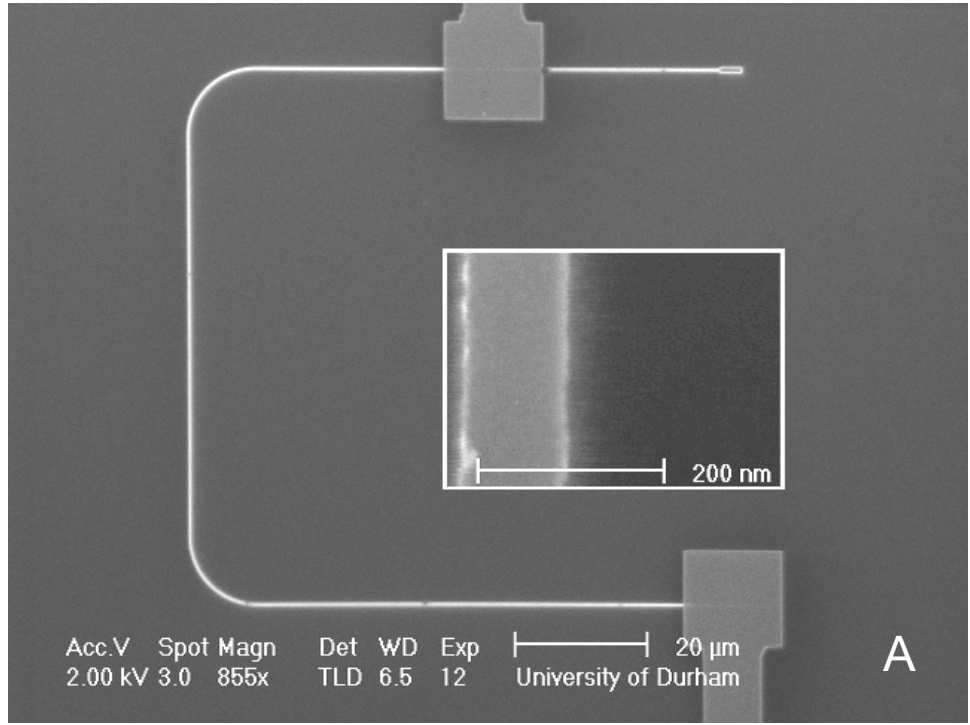

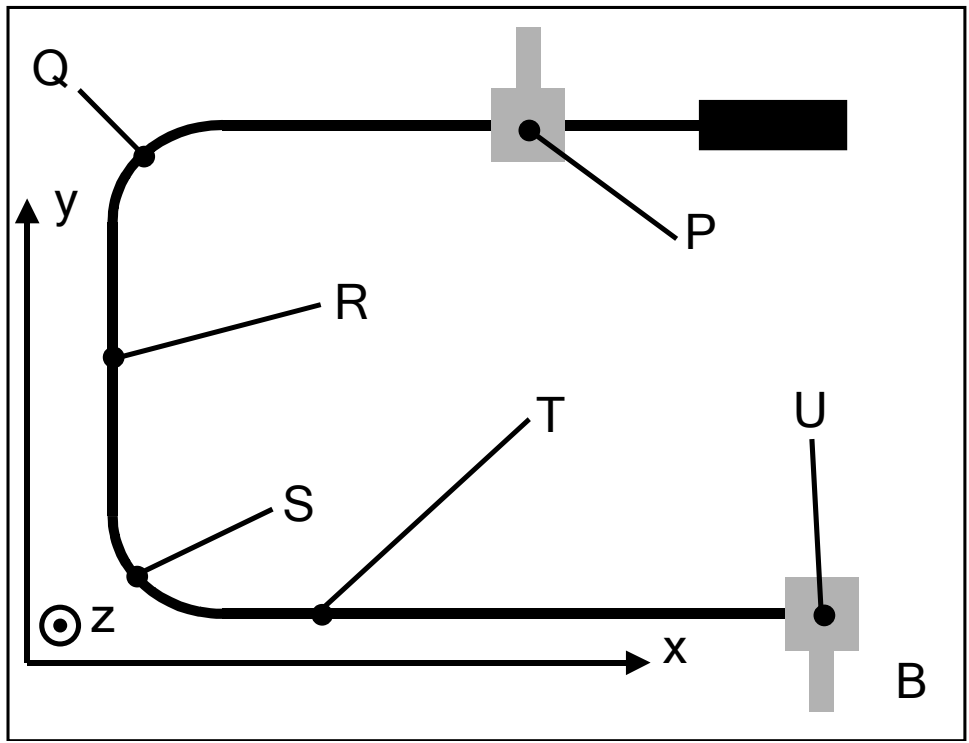

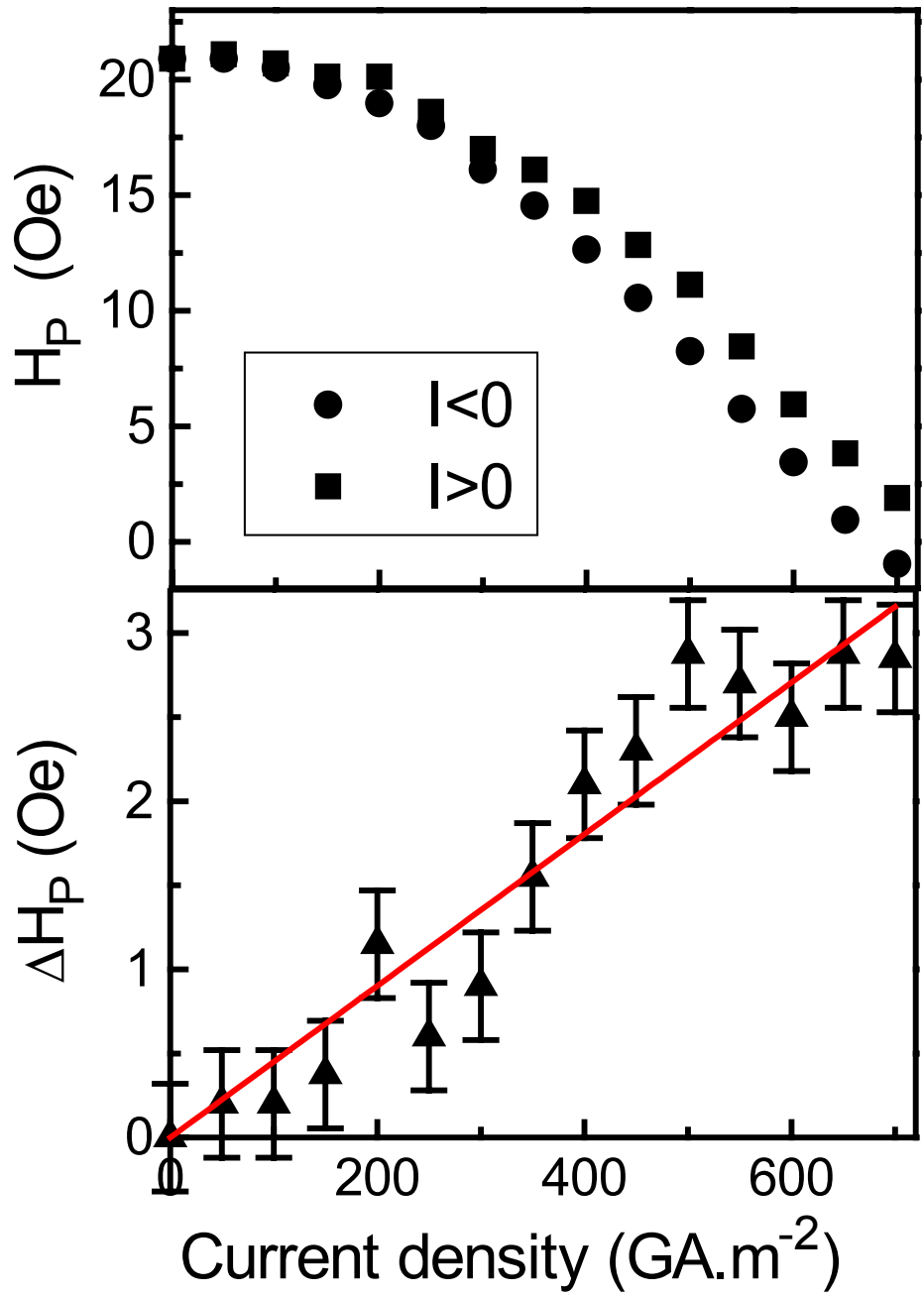

**Figure 3**

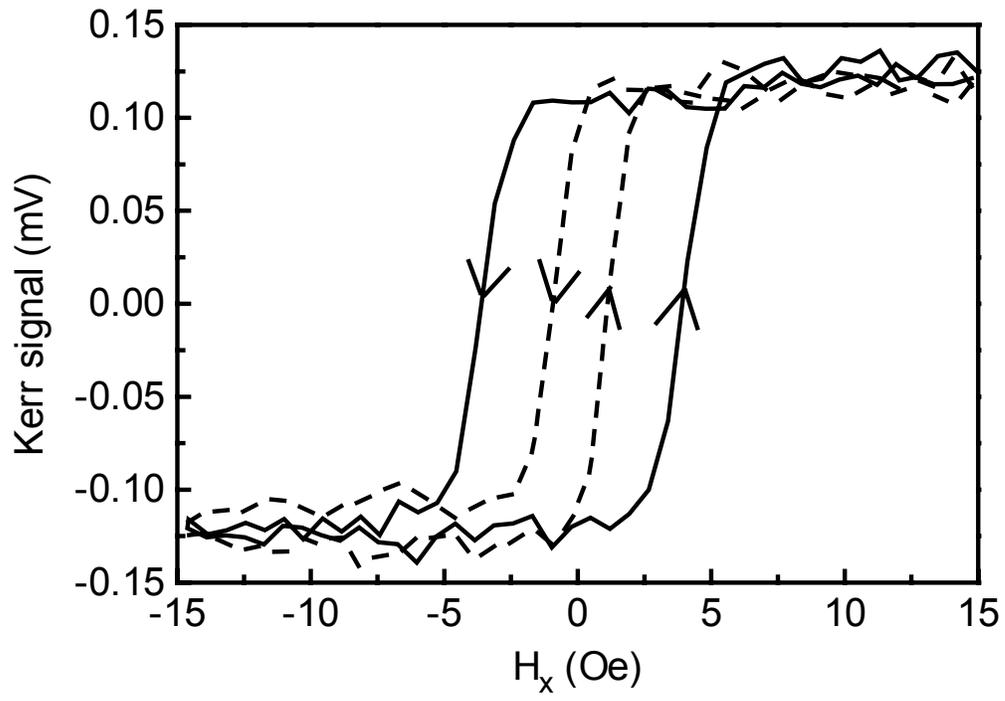

**Figure 4**

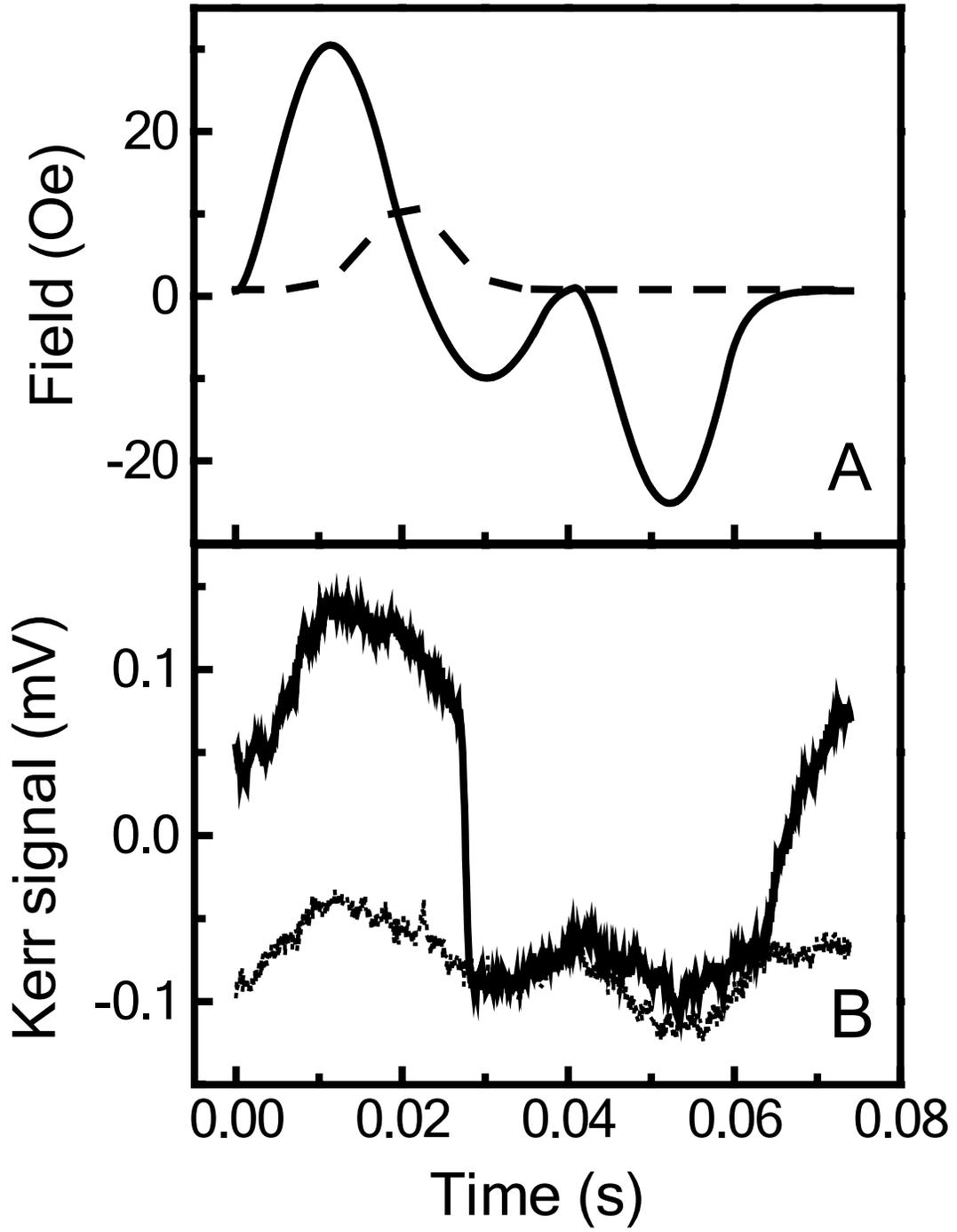